# FAST *K*-MEANS ALGORITHM CLUSTERING

Raied Salman[1], Vojislav Kecman, Qi Li, Robert Strack and Erick Test

[1]Department of Computer Science, Virginia Commonwealth University, Virginia, 601 West Main Street, Richmond, VA 23284-3068, USA
{salmanr, vkecman, liq, strackr, estest}@vcu.edu

## ABSTRACT

*k-means has recently been recognized as one of the best algorithms for clustering unsupervised data. Since k-means depends mainly on distance calculation between all data points and the centers, the time cost will be high when the size of the dataset is large (for example more than 500millions of points). We propose a two stage algorithm to reduce the time cost of distance calculation for huge datasets. The first stage is a fast distance calculation using only a small portion of the data to produce the best possible location of the centers. The second stage is a slow distance calculation in which the initial centers used are taken from the first stage. The fast and slow stages represent the speed of the movement of the centers. In the slow stage, the whole dataset can be used to get the exact location of the centers. The time cost of the distance calculation for the fast stage is very low due to the small size of the training data chosen. The time cost of the distance calculation for the slow stage is also minimized due to small number of iterations. Different initial locations of the clusters have been used during the test of the proposed algorithms. For large datasets, experiments show that the 2-stage clustering method achieves better speed-up (1-9 times).*

## 1. INTRODUCTION

No theoretical research work available on the running time was required for the $k$-means to achieve its goals as mentioned by [1]. They researched the worst-case running time scenario as superpolynomial by improving the lower bound from $\Omega(n)$ iterations to $2^{\Omega(\sqrt{n})}$. [9] has developed another method to reduce the number of iterations but it was not as fine-tuned as [1]. On the other hand [4] have proved that the number of iterations required by $k$-means is much less than the number of points. Moreover, [5] were unable to bound the running time of $k$-means, but they proved that for every reclassified point one iteration is required. Then after $O(kn^2\Delta^2)$ iterations the convergence will be guaranteed.

A group of researchers worked on choosing the best centers to avoid the problems of $k$-Means of either obtaining the non-optimal solutions or empty clusters generations. [3] worked on modifying the $k$-means to avoid the empty clusters. They moved the center of every cluster into new locations to ensure that there will be no empty clusters. The comparison between their modified $k$-means and the original $k$-means show that the number of iterations is higher with the modified $k$-means method. In case of the numerical examples which produce empty clusters, the proposed method cannot be compared with any other method since there is no modified $k$-means algorithm available to avoid the empty clusters. [6] on the other hand developed a procedure in which the centers have to pass a refinement stage to generate good starting points. [7] used genetically guided $k$-means where the possibility of empty clusters will be treated in the mutation stage. Another method of center initializing based on values of attributes of the dataset is proposed by [8]. The later proposed method creates a complex procedure which leads to be computationally expensive.

[2] on the other hand, developed a method to avoid unnecessary distance calculations by applying the triangle inequality in two different ways, and by keeping track of lower and upper bounds for distances between points and centers. This method is effective when the dimension is more than 1000 and also when the clusters are more than 20. They claimed that their method is many times faster than normal $k$-means method. In their method the number of distance calculations is $N$ instead of $KQN$ where $N$ is the number of points and $KQ$ are the number of clusters and the number of iterations respectively. [9] In contrast, Hodgson used different triangular equality to achieve the goal, in which they reduced the number of distance calculations.

## 2. THEORETICAL BACKGROUND AND THE PROPOSED METHOD

Simple modifications of $k$-means clustering method have been proposed. The theoretical background of the proposed method is described below:

The main idea behind $k$-means is to calculate the distance between the data point and the centers using the following formula:

$$d(x_k, x_c) = [(x_k - x_{ci})^T (x_k - x_{ci})]^{1/2} \qquad (1)$$

Where $d$ the Euclidean distances between the data point $x_k$ at the cluster $k$ and the initial centers are $x_{ci}$

The points in one cluster are defined as:

$x_l$ for $l = 1, 2, \ldots, n$ regarded as one cluster and $n$ is the total number of points in that cluster.

The $x_{ci}$ chosen randomly either from the dataset or arbitrarily. In our method we have used the random selection of the centers from the dataset to avoid wasting one more calculation (iteration). Any $k$-means clustering method depends on the number of clusters set at the beginning. There is no guarantee that the centers will move or converge to the mean points of the average of the cluster. This is one of the drawbacks of $k$-means. Also there is no guarantee that the convergence will happen to the local mean.

Assume that $A_n$ is the set of $i$ clusters to minimize the criteria $J(.;P)$ so that $x_{ci}$ converges to $x_i$ (the cluster centers):

$$A_n = \{x_{c1}, x_{c2}, \ldots, x_{ci}\} \qquad (2)$$

where $\qquad J(x_1, x_2, \ldots, x_i; P) = P(\min_i |x_k - x_i|^2) \qquad (3)$

where $P$ is the probability distribution over the Euclidean space.

If the $S_n$ represents the entire dataset then the objective is to find a subset $S_s$ of $S_n$ such that $P(S_s) \leq P(S_n)$

We assume that the data with one center is a stationary random sequence satisfying the following cumulative distribution sequence:

$$F_{X_n, X_{n+1}, \ldots, X_N}(x_n, x_{n+1}, \ldots, x_N) = F_{X_{n+k}, X_{n+1+k}, \ldots, X_{N+k}}(x_n, x_{n+1}, \ldots, x_N) \qquad (4)$$

then the above sequence has one mean:

$$E(X) = c \qquad (5)$$

The process of clustering is equivalent to minimizing the Within-Cluster Sum of Squares for the, so called, fast stage:

$$\min_S \sum_{i=1}^{c_i} \sum_{x_j \in S_i} \|x_j - \mu_{fa}\| \qquad (6)$$

and for the so called, slow stage, as follows:

$$\min_S \sum_{i=1}^{c_i} \sum_{x_j \in S_i} \|x_j - c_{sl}\| \qquad (7)$$

where $c$ are the centers of the clusters which are equals to the centers of the previous stage.
The within cluster sum of squares is divided into two parts corresponding to the fast and the slow stages of the clustering:

$$WSCC = \int_0^{c_i} min(\|x-c\|, \|x-\bar{c}\|)dx + \int_{c_i}^1 min(\|x-c\|, \|x-\bar{c}\|)dx \qquad (8)$$

The centers of the slow stage start with $c_i$

The following algorithm describes briefly the proposed procedure:

---
**Algorithm:**
**Input:**    $S_n, c, per, J_f, J_s$
**Output:**    $S_n$ with clusters
$S_n = \%per$ of $S_n$
Select $X_c$ from $S_n$ randomly
**While** $J_f \leq \mu_{fa}$
    **For** $i \leftarrow 1$ to $n_s$
        Calculate the modified distance
        $d(x_k, x_c) = [(x_k - x_{ci})^T(x_k - x_{ci})]$
        Find minimum of $d$
        Assign the cluster number to point $X_i$
    **End for**
    Calculate $J_f$
**End while**
Calculate the average of the calculated clusters to find new centers $X_c$
Use the whole dataset $S_n$
**While** $J_s \leq \mu_{sl}$
    **For** $i \leftarrow 1$ to $n$
        Calculate the modified distance
        $d(x_k, x_c) = [(x_k - x_{ci})^T(x_k - x_{ci})]$
        Find minimum of $d$
        Assign the cluster number to point $X_i$
    **End for**
    Calculate $J_s$
**End while**

---

## 3. PERFORMANCE IMPROVEMENT

The complexity of the $k$-means is $O(KQN)$ where $K$ is the number of clusters, $Q$ is the number of iteration required to get to the stopping criteria and $N$ is the input patterns. For example if the data size is 1000 points, 4 clusters and it require 20 iterations to get the optimal locations of the centers. Then, $O(80,000)$ is the time complexity. The time complexity in the proposed method has two folds, first is time complexity of the fast stage of clustering:
$O(KQ_fN_f)$ where $N_f$ is the number of data for the fast stage and $Q_f$ is the iterations during the fast stage only. The second part of the time complexity is calculated according to the slow stage of clustering: $O(KQ_sN)$ where $Q_s$ is the number of iterations during the slow stage. Assume that $N_f = 100$ and $Q_s = 3$ then the total time complexity is:
$O(KQ_fN_f) + O(KQ_sN) = O(7200) + O(8000) = O(15200)$

This will represent a reduction in the calculation time for the clustering of more than 5 times. However, if the data is bigger than the previous figure then the time reduction will be higher. For example if the data is 1Million the reduction will be approximately 10 times. This is quite well illustrated in the following diagram:

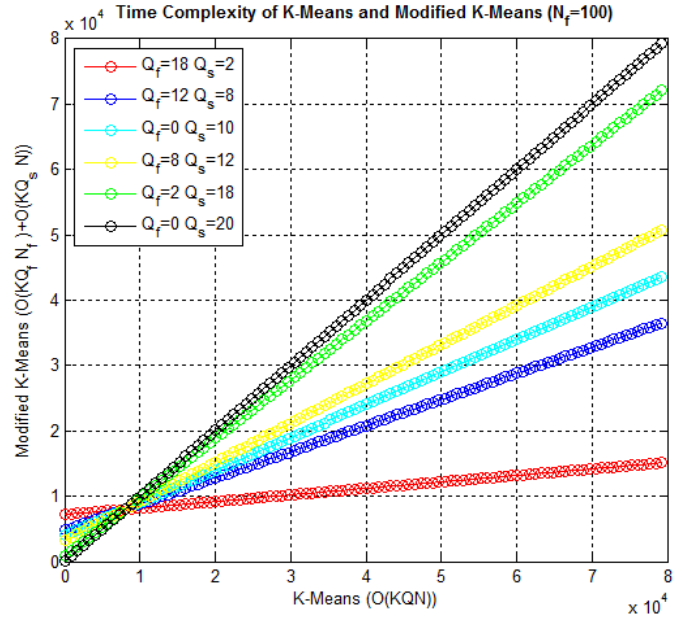

Figure 1. Complexity measure of the k-means and the modified k-means with 100 samples

The $Q_f$ and $Q_s$ are the fast iterations and the slow iterations of the modified $k$-means. Respectively. The black graph in Fig. 1 is the time complexity of the normal $k$-means. Other graphs represent the complexity of the modified $k$-means. Therefore the higher the value of $Q_s$ the more the graphs will approach the normal $k$-means properties. From the above graph it can be concluded that the lower values of $Q_f$ the less time required to achieve total clustering. The more iterations, for the fast stage, the faster the algorithm works. However, the catch here is we cannot go very low with $Q_f$ as the time of the clustering will approach the normal $k$-means. In other words the clustering procedure will produce blank clusters. The proper percentage would be 10% - 20%. The set up of the parameters of the red graph of the above diagram has a complexity of almost 5 times less than the normal $k$-means clustering. In the case of using higher number of data for the fast stage clustering, $N_f = 500$, the complexity results will be skewed upwards as shown below:

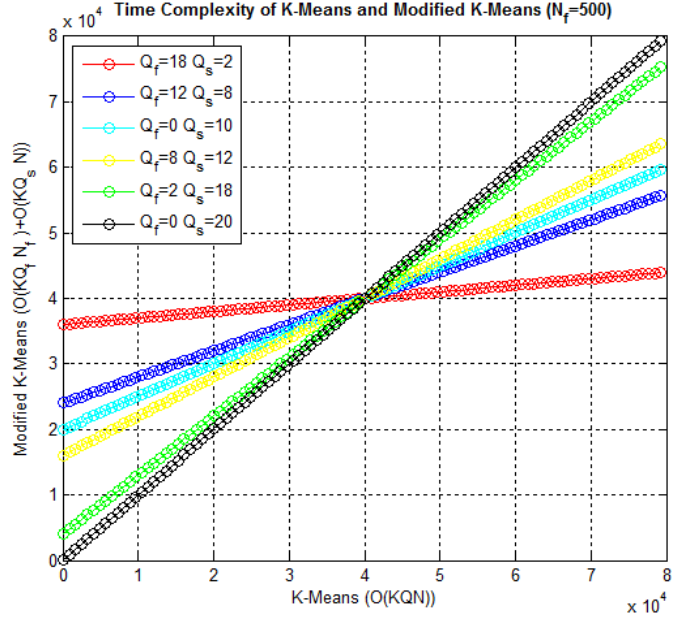

Figure 2. Complexity measure of the k-means and the modified k-means with 500 samples

The set up of the parameters of the red graph of the above diagram has complexity less than 2 times than the normal $k$-means clustering. This indicates that the more the data chosen for the fast stage of clustering the less advantages of this method.

## 4. ANALYSIS OF THE ALGORITHM

To investigate the suitability of the proposed method we run the simulation for many different parameters. The parameters which have to be adjusted to get the best speed up values are: The data size, the dimension of the data, the number of clusters, the stopping criteria of the first clustering stage, the stopping criteria for the second clustering stage and the percentage of the data used for the 2-stage clustering. Another important consideration which has been taken into consideration is the use of the same program for running the normal k-means clustering and the 2-stage clustering after feeding it with different centers and the use of part of the data. Furthermore, the same computer has been used for running all simulations to avoid the discrepancy of the computer performance. The computer used is an Alienware with i7 CPU and 6 GB RAM. For the validation of the proposed method, random and uniform clusters were created. Two examples are shown below to validate the application of the 2-stage k-means algorithm.

### 4.1. Validation of the 2-stage k-means clustering algorithm

Assume that $X \subset \mathbb{R}^d$ is a finite set of $n$ points and $d$ is the dimension of the data (features). The number of clusters is $k$ which is an integer >1 since we consider that the data belong to more than one cluster. The clustering procedure is to find $S = (S_1, \ldots, S_k)$ groups in which the data is divided into the $S$ clusters without assigning one point into two or more clusters. Every cluster has one center such that all centers are $C = (C_1, \ldots, C_k)$.

We also assumed that there exit subset such that it satisfied the condition that all the points in the subset cover all clusters.

The k-means clustering algorithm used in this work is also known as Lloyd's method; see [12].

*Theorem*

Given a data $X_f$ as subset of $X$ with $S^f = \left(S_1^f, \ldots, S_k^f\right)$ clusters and centers $C = (C_1, \ldots, C_k)$, then applying one step of k-means would result in shifting the centers to the new locations $C^f = \left(C_1^f, \ldots, C_k^f\right)$ such that $C_1^f\left(S_1^f\right), \ldots, C_k^f\left(S_k^f\right)$, so if $X_f \Rightarrow X$ then

$$C_1^f\left(S_1^f\right), \ldots, C_k^f\left(S_k^f\right) \Rightarrow C_1(S_1), \ldots, C_k(S_k)$$

Proof of Theorem

We assume that the whole data set is $X$ of size $n$ and subset of it is $X_f$ with $\leq n$, where $f$ is an integer representing the number of data in the fast stage. Also all clusters $S = (S_1, \ldots, S_k)$ exist in $X_f$, in other words there is no empty cluster. This means that $mean(C_i^f) \neq mean(C_i)$ if $f \neq n$. However, if $f \Rightarrow n$ then $mean(C_i^f) \Rightarrow mean(C_i)$

This completes the proof

*Corollary*

Continuing from the above Theorem, we can conclude that if the centers $C_1^f\left(S_1^f\right), \ldots, C_k^f\left(S_k^f\right)$ are produced from the fast stage, then it can be used for another k-means who uses the whole data $X$ without the loss of generality of k-means clustering. This is what we call slow stage clustering.

Since the centers produced from the fast stage $C_1^f\left(S_1^f\right), \ldots, C_k^f\left(S_k^f\right)$ are located on the path of convergence of the centers for each cluster, then using these centers for clustering with whole data will be a valid option. This ends the proof of the corollary. Figure3 shows that always the fast stage cluster centers converges to the cluster centers, especially when the number of data is high.

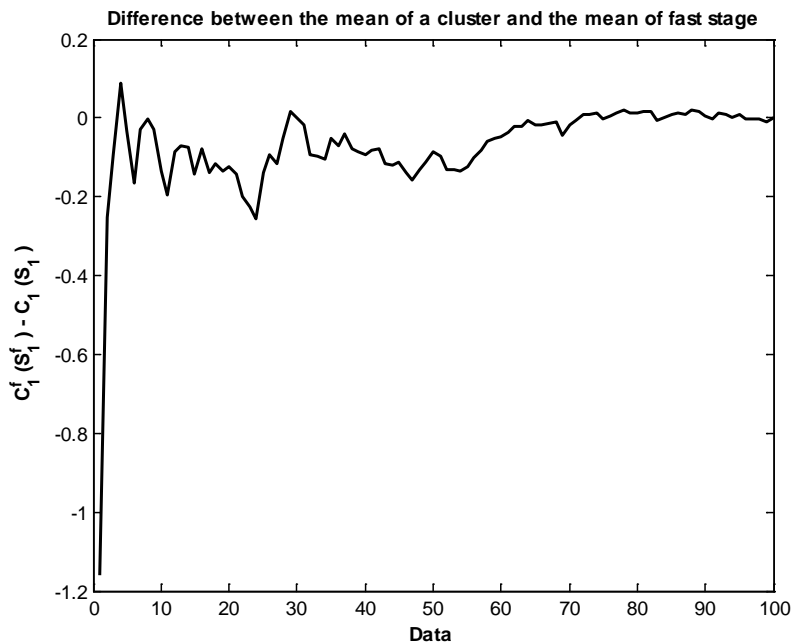

Figure 3. Convergence of the cluster center with the fast stage cluster center

## 5. NUMERICAL EXAMPLES

Two examples presented here to validate the proposed method.
A data set with 800 samples and 2-dimension (3 clusters) is used. The following figures show the movement of one of the centers and the two stage clustering.

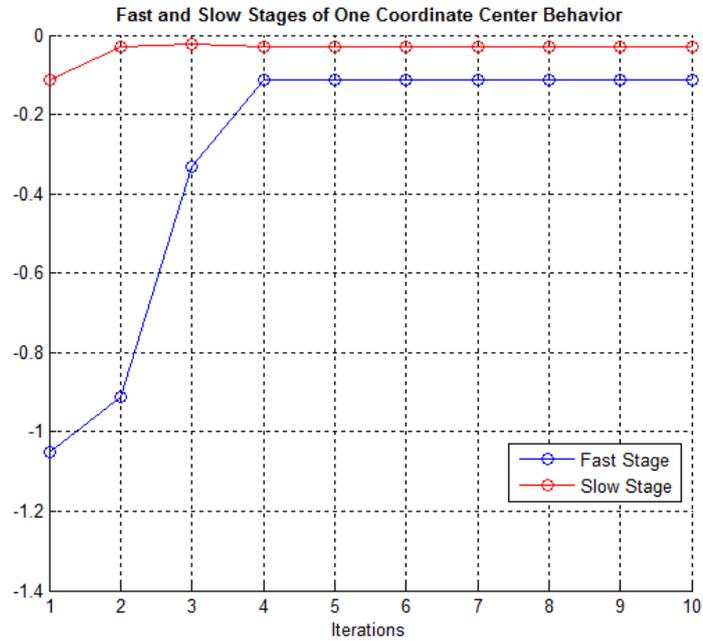

Figure 4. Fast and Slow stages of the movement of one coordinate during the clustering

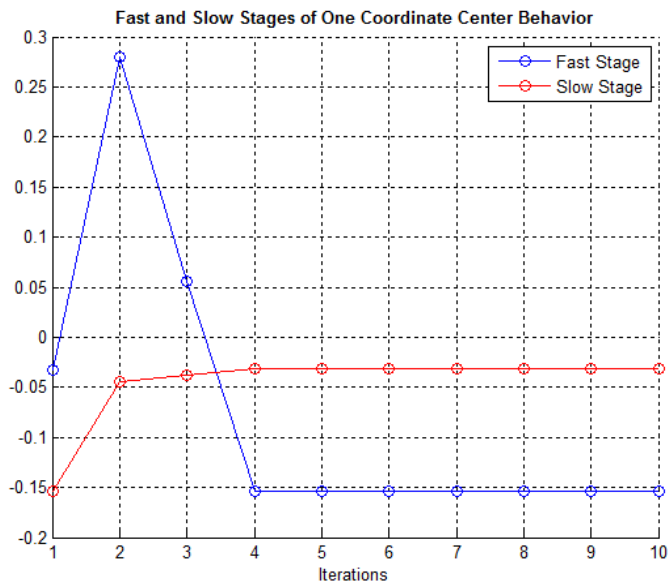

Figure 5. Fast and Slow stages of the movement of the second coordinate during the clustering

From Figs. 3 and 4 it is very clear that the approach of the red line (slow stage coordinate of one center) is very smooth comparing with the other fast stage coordinate movements. The first value of the red (slow) graph is the same as the last value of the blue (fast) graph. The number of iterations is higher than is required but this is only for clarification. The number of iterations required for the fast stage will of course be higher than the slow stage scheme.

Moreover, as you can see from the above graph, the coordinates have not been changed a lot. This means that the $k$-means algorithm does not need to run many times since we reached the correct accuracy.

Another presentation of the effectiveness of the method is the movements of the three centers as shown in figures 6-9.:

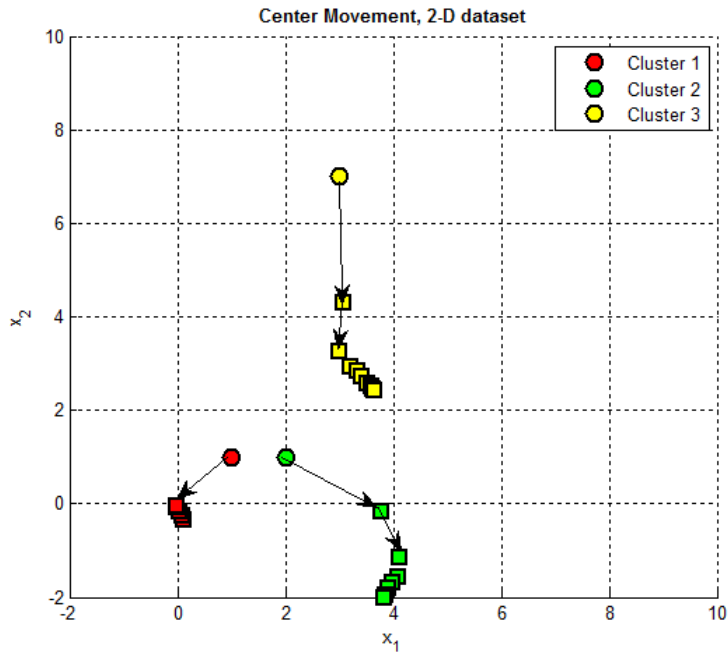

Figure 6. Three center movement during the fast stage clustering.

A more detailed description is shown in the following figures in which the fast stage shows the squares and the slow stage shows the diamond symbol:

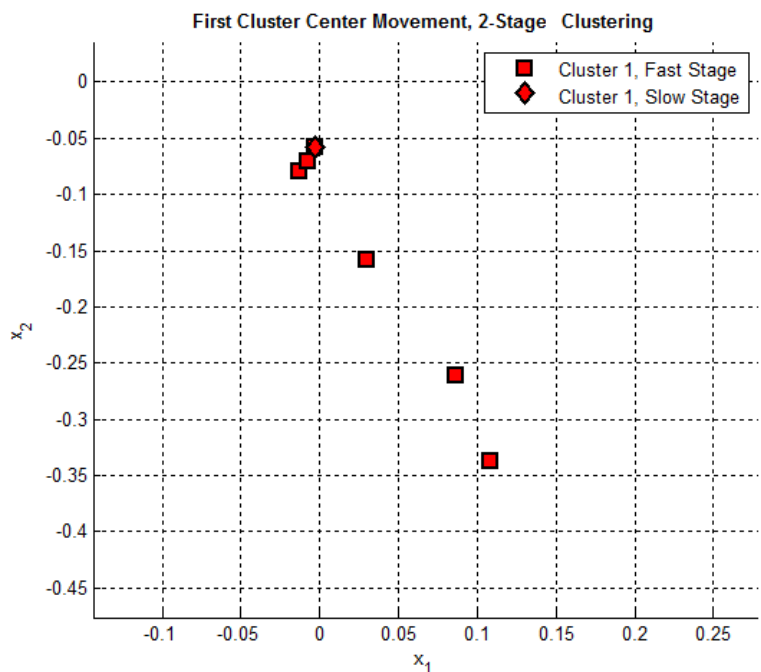

Figure 7. Fast and slow stages of the first cluster center movements.

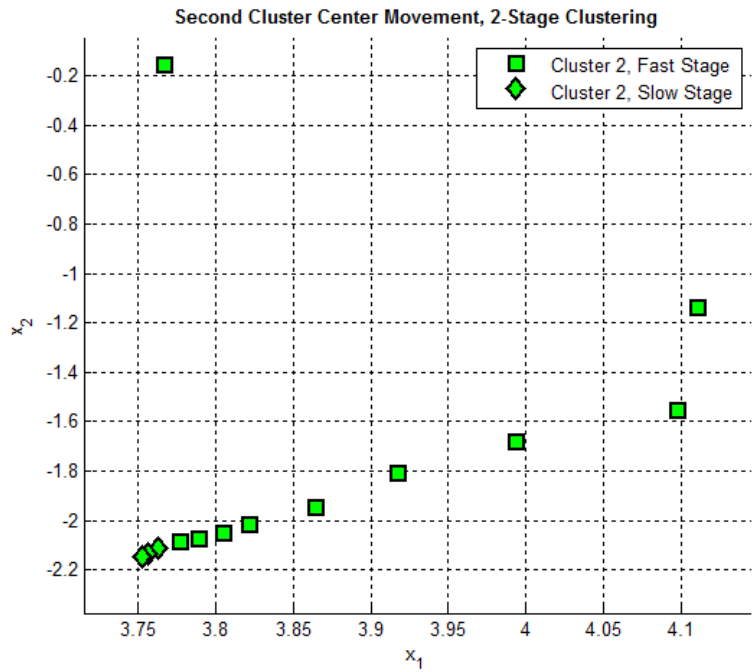

Figure 8. Fast and slow stages of the second cluster center movements.

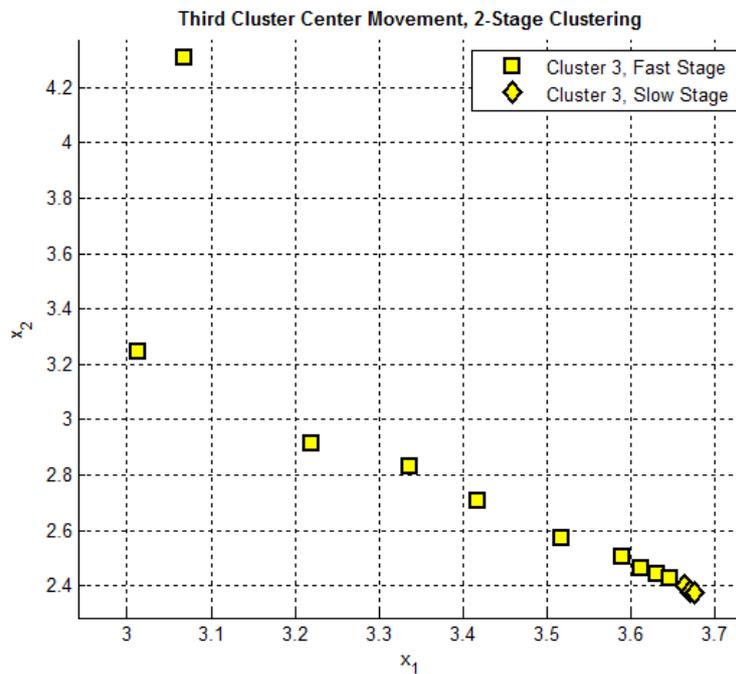

Figure 9. Fast and slow stages of the third cluster center movements.

As can be seen from the above diagrams, that the centers have moved many steps during the fast stage, this has been achieved in fast response. The diamond shapes shows the slow stage of iteration. The number of iterations of the slow stage is much less than the fast stage. Also the movements of the centers are very small. In this case the required calculation would be reduced from many steps to only couple of full step (for all dataset). This of course will save some time and reduce expenses.

To be more specific about the proposed method Table 1 shows the creation of clusters in different iterations for three dimensional data.

Table 1. Distribution of points and centers during the fast and the slow stages of clustering

| Stages | Iter No. | Clusters | Old Centers | | | New Centers | | | Points in Clusters | Points |
|---|---|---|---|---|---|---|---|---|---|---|
| Fast | 1 | C1 | 8 | 4 | 4 | 4.867 | 3.267 | 1.567 | 30,38,44 | 15 |
| | | C2 | 4 | 4 | 4 | 6.16 | 2.85 | 4.68 | 53,58,72 86,88,93 113,114 138,145 | |
| | 2 | C1 | 4.867 | 3.267 | 1.567 | 4.867 | 3.267 | 1.567 | 30,38,44 | |
| | | C2 | 6.16 | 2.85 | 4.68 | 6.16 | 2.85 | 4.68 | 53,58,72 86,88,93 113,114 138,145 | |
| Slow | 1 | C1 | 4.867 | 3.267 | 1.567 | 5.015 | 3.318 | 1.636 | 1-50,58 61,82,94 99 | 150 |
| | | C2 | 6.16 | 2.85 | 4.68 | 6.323 | 2.901 | 4.987 | 51-57,59-60 62-81 83,93 95-98 100-150 | |
| | 2 | C1 | 5.015 | 3.318 | 1.636 | 5.06 | 3.226 | 1.897 | 1-50,54,58 60-61,70 80-82,90 94,99,107 | |
| | | C2 | 6.323 | 2.901 | 4.987 | 6.396 | 2.933 | 5.071 | 51-53,55-57 59,62-69 71-79 83-89 91-93 95-98 100-106 108-150 | |
| | 3 | C1 | 5.06 | 3.226 | 1.897 | 5.083 | 3.205 | 1.956 | 1-50,54,58 60-61 63 65,70,80-82 90,94 99,107 | |
| | | C2 | 6.396 | 2.933 | 5.071 | 6.409 | 2.942 | 5.1 | 51-53,55-57 59,62,64 66-69,71-79 83-89,91-93 95-98 100-106 108-150 | |
| | 4 | C1 | 5.083 | 3.205 | 1.956 | 5.083 | 3.205 | 1.956 | 1-50,54,58 60-61,63,65 70,80-82,90 94,99,107 | |
| | | C2 | 6.409 | 2.942 | 5.1 | 6.409 | 2.942 | 5.1 | 51-53,55-57 59,62 64,66-69 71-79 83-89 91-93,95-98 100-106 108-150 | |

The two stages are clearly indicating the formation of clusters at each stage. The centers of the slow stage are the same as the centers of the end of the fast stage. The fast stage has taken only 10% of the whole dataset. Although the number of iterations of the slow stage is 4, in bigger dataset this can be reduced by increasing the iterations of the fast stage.

One data set from the UCI repository has been chosen to insure the applicability of the proposed method in [10]. The dataset is "Synthetic Control Chart Time Series" with 600 points, 60-

dimesion and 6 clusters, [11]. We have chosen one center and one coordinate of the 60 dimension to clarify the movements of the centers as indicated by Figure 10.

The movement of the one coordinate is shown in Figure 8 for almost 300 iterations. This movement of the centers is measured as the difference between the mean of the cluster and the centers. This difference changes all the time during the iterations.

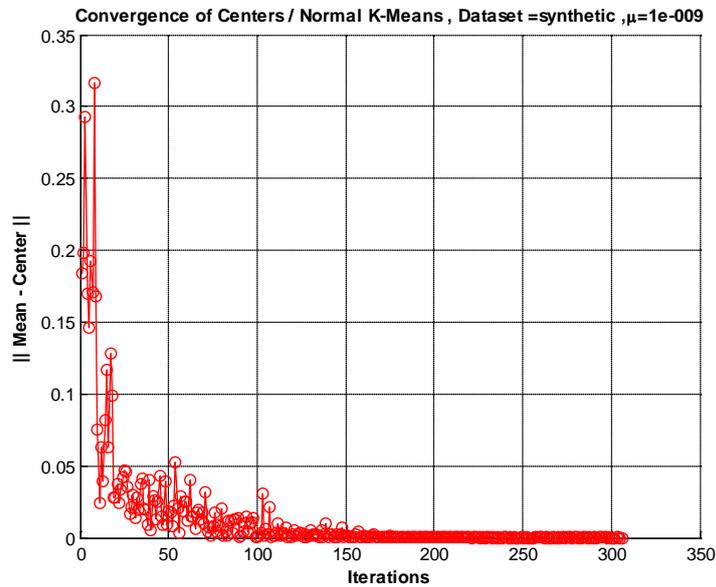

Figure 10. Center convergence of one coordinate for one center of the Synthetic dataset using the normal k-means algorithm

The use of the 2-stage k-means clustering algorithm for the Synthetic data set to calculate the center convergence is shown in Figure 9. The stopping criterion for fast stage is $10^{-3}$ and for the slow stage is $10^{-6}$ as the normal k-means method. Figure 11, however shows the fast convergence (blue) and slow convergence (red). 10% of the data has been used for the fast stage. The fluctuation of the centers during the slow stage is much less than the fluctuation of the centers in the normal k-means of Figure 10.

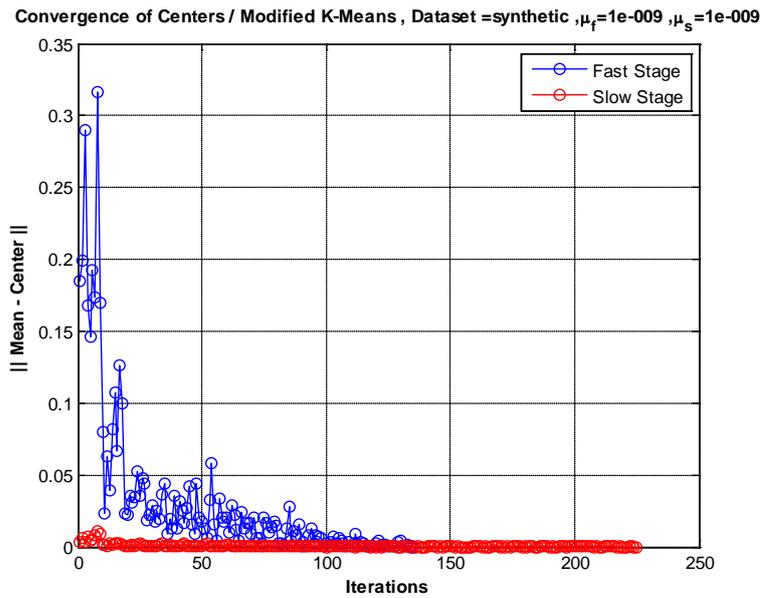

Figure 11. Center convergence of one coordinate for one center of the Synthetic dataset using the fast (blue) and slow (red) k-means algorithm

## 6. SPEED UP ANALYSIS

The speed of the normal $k$-means is shown in blue while the speed of the modified $k$-means is shown in red. Two different computers were used of 32bit and 64bit Operating Systems. Regardless, of the speed of the computer used the validation of the modified $k$-means always consistent as indicated by Fig. 9. The data used for the fast stage clustering is only $10\%$ of the whole data which is randomly selected. The dataset used in this example is "Synthetic" which is $100,000$ samples with 10 dimensions. The speed of the modified $k$-means is almost twice the speed of normal $k$-means. This is due to the fact that 2-stage $k$-means clustering uses less full data iterations. The speed up is very clear in the high accuracy when the required µ is $10^{-4}$ or less, where µ is the stopping criteria or the required accuracy. This is always important when you try to find good clustering results.

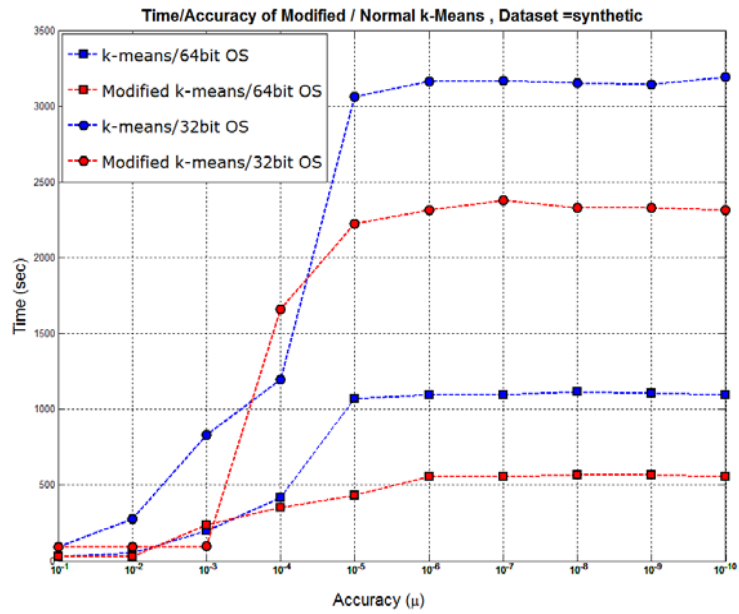

Figure 12. Comparison in the speed of the modified $k$-means and normal $k$-means with different computers

The speed up of the modified $k$-means comparing with the normal $k$-means is varying according to the accuracy. For the lower range of accuracy the speed up of clustering is ranges from (1-9) times. This would reduced for the higher accuracy for example from $10^{-4}$ to $10^{-10}$. Figure 9 shows clearly that the speed up is settled for the higher accuracy within 2 times. On the other hand the range of the random data selected to archive the fast clustering is also fluctuating. The best range is between 10%-20%. In the normal situation we require a good accuracy for the clustering to archive the full clustering to all data. This would be between 10%-20% of the data and accuracy between $10^{-4}$ to $10^{-10}$ as shown in Table 2.

Table 2. Speed up of clustering with the modified k-means using different dataset sample

| | | percentage of the total data | | | | |
|---|---|---|---|---|---|---|
| | | 10% | 15% | 20% | 30% | 40% |
| Accuracy | $10^{-1}$ | 1.9 | 1.8 | 1.8 | 1.7 | 1.5 |
| | $10^{-2}$ | 3.8 | 3.5 | 3.4 | 3 | 2.5 |
| | $10^{-3}$ | 4.7 | 8.9 | 3.1 | 7 | 4.3 |
| | $10^{-4}$ | 1 | 1.7 | 1.1 | 3 | 8.5 |
| | $10^{-5}$ | 2.9 | 1.6 | 2.2 | 2.1 | 2.4 |
| | $10^{-6}$ | 2 | 1.9 | 2.6 | 2.3 | 2.4 |
| | $10^{-7}$ | 2 | 1.4 | 2.4 | 2.3 | 1.6 |
| | $10^{-8}$ | 2 | 1.4 | 2.4 | 2.3 | 1.6 |
| | $10^{-9}$ | 2 | 1.4 | 2.4 | 2.3 | 1.6 |
| | $10^{-10}$ | 2 | 1.4 | 2.4 | 2.3 | 1.6 |

The proper range of the sample data is between 10%-20%. Carrying out the required time for running the normal $k$-means and the modified $k$-means for 9 different data samples shows that the best range is 10%-20% to get less time in the calculation of the two algorithms as shown in Table 3.

Table 3. Calculation time for normal kmeans and modified kmeans

|  | Fast + Slow k-means (sec) | Normal k-means (sec) |
|---|---|---|
| 10% | 4.81 | 14.3 |
| 20% | 9.93 | 14.3 |
| 30% | 14.94 | 14.3 |
| 40% | 19.95 | 14.3 |
| 50% | 25.17 | 14.3 |
| 60% | 30.45 | 14.3 |
| 70% | 36.62 | 14.3 |
| 80% | 42.01 | 14.3 |
| 90% | 47.66 | 14.3 |

(Percentage of the data)

## 6.1. The Speed of the 2-stage k-means and normal k-means algorithms

Bigger data sets were used to find out the time required to achieve the same results for both methods, the 2-stage and the normal k-means algorithms. As indicated previously that we used the same machine to run all examples. The data used is: 1M, 2M, 3M, 4M, 5M, 6M, 7M, 8M, 9M, 10M with 12-dimensions. We kept the stopping criteria the same for both methods as indicated earlier. Since the data used in this step is much higher than the ones used previously, we only used a small part of the data (1%) for the calculation of the fast stage. Obviously, this small data will be chosen at random, again to insure the proper distribution amongst the whole datasets. The red circles in Figure 13 represent the normal k-means clustering method. While the blue circles characterize the 2-stage k-means algorithm. As you can see, the difference between the two speeds is higher for the higher datasets. The time required for the normal k-means clustering using the 10M dataset is almost 2200 seconds, while the time for the 2-stage method is 640 seconds. Such reduction in time consumed for calculation is very useful and cost effective.

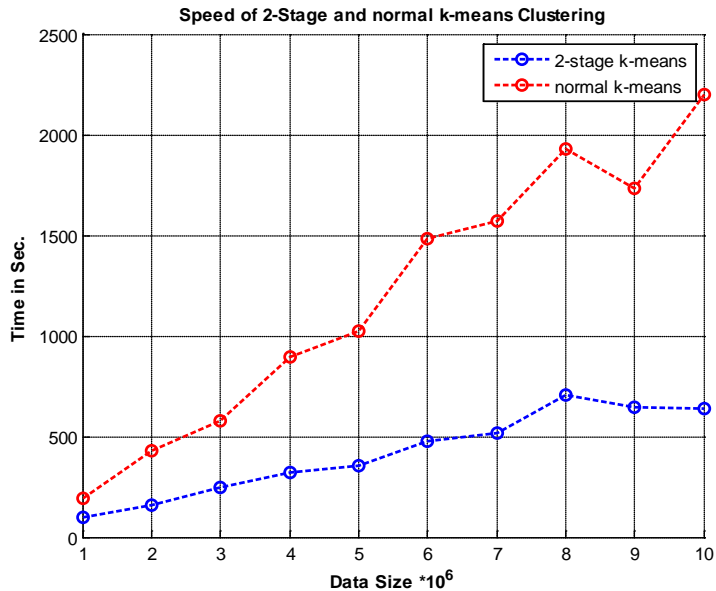

Figure 13. The speed of the normal k-means and the 2-stage k-means clustering algorithms

The limitation of choosing bigger data sets prevented us from validating the method for data bigger than 10M points. This would be interesting finding for anybody who can run the above method for bigger data sets.

# 7. CONCLUSIONS

A simple proposal for achieving high speed of $k$-means clustering for ultra datasets has been presented in this paper. The idea has two folds. The first is the fast calculation of the new centers of the $k$-means clustering method. A small part of the data will be used in this stage to get the final destination of the centers. This of course will be achieved in high speed. The second part is the slow stage in which the $k$-means will start from well positioned centers. This stage may take a couple of iterations to achieve the final clustering. The whole dataset will be used for the second stage.

In normal $k$-means algorithms, if the initial centers are located exactly at the means of the clusters of the data, then the algorithm requires only one step to assign the individual clusters to each data point. In our modified $k$-means we are trying to get to the stage of moving any initial centers to a location which is either that of the means or near them. The big gap between these locations will decide how many times the normal $k$-means is required to run to assign all data to their clusters. Our algorithm will quickly move the centers to locations which are near the means. Future work is required to find out the effect of different locations of the clusters on the speed up.

# REFERENCES

[1] Arhter, D. and Vassilvitskii, S.: How Slow is the *k*Means Method? SCG'06, Sedona, Arizona, USA. (2006).

[2] Elkan, C.: Using the Triangle Inequality to Accelerate K –Means. Proceedings of the Twentieth International Conference on Machine Learning (ICML-2003), Washington DC, (2003).

[3] Pakhira, Malay K.: A Modified *k*-means Algorithm to Avoid Empty Clusters. International Journal of Recent Trends in Engineering, Vol 1, No. 1, May (2009).

[4] Dude, R. O., Hart, P. E., and Pattern, D. G.: Classification. Wiley-Interscience Publication, (2000).

[5] Har-Peled, S. and Sadri, B.: How fast is the k-means method? Algorithmica, 41(3):185–202, (2005).

[6] Bradley, P. S. and Fayyad, U. M.: Refining Initial Points for Kmeans Clustering. Technical Report of Microsoft Research Center, Redmond,California, USA, (1998).

[7] Wu, F. X.: Genetic weighted k-means algorithm for clustering large-scale gene expression data. BMC Bioinformatics, vol. 9, (2008).

[8] Khan, S. S. and Ahmed, A.: Cluster center initialization for Kmeans algorithm. Pattern Recognition Letters, vol. 25, no. 11, pp. 1293-1302, (2004).

[9] Hodgson, M. E.: Reducing computational requirements of the minimum-distance classifier. Remote Sensing of Environments. *25*, 117–128, (1988).

[10] R. Salman, V. Kecman, Q. Li, R. Strack and E. Test, "Two-Stage Clustering with k-means Algorithm," WIMO 2011 Conference, Ankara, Turky, June 2011 (in press).

[11] R. J. Alcock and Y. Manolopoulos, "Time-Series Similarity Queries Employing a Feature-Based Approach," 7th Hellenic Conference on Informatics. August 27-29. Ioannina, Greece 1999.

[12] S. Lloyd, "Least squares quantization in pcm," IEEE Transactions on Information Theory, 28:129–137, 1982.

## Authors

Raied Salman received the B.S. degree (with high distinction) in electrical engineering and M.S. degree in Computer Control from the University of Technology, Baghdad, Iraq in 1976 and 1978, respectively. He also received the Ph.D. in Electrical Engineering from Brunel University, England, UK in 1989. He is currently a Ph.D. candidate in the Department of Computer Science at Virginia Commonwealth University, Richmond, VA. His research interests include Machine Learning and data mining for large datasets.

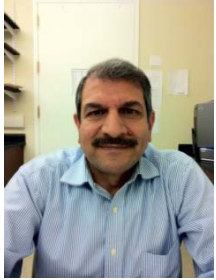

Vojislav Kecman is with VCU, Dept. of CS, Richmond, VA, USA, working in the fields of machine learning by both support vector machines (SVMs) and neural networks, as well as by local approaches such as Adaptive Local Hyperplane (ALH) and Local SVMs, in different regression (function approximation) and pattern recognition (classification, decision making) tasks. He was a Fulbright Professor at M.I.T., Cambridge, MA, a Konrad Zuse Professor at FH Heilbronn, DFG Scientist at TU Darmstadt, a Research Fellow at Drexel University, Philadelphia, PA, and at Stuttgart University. Dr. Kecman authored several books on ML (see www.supportvector.ws and www.learning-from-data.com).

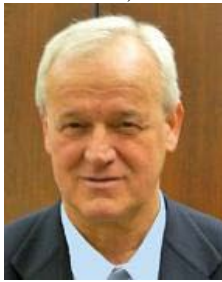

Qi Li received his B.S. degree in Electronic Engineering from Beijing University of Posts and Telecommunications, Beijing, China, in 2007 and M.S. degree in Computer Science from Virginia Commonwealth University, Richmond, United States, in 2008. He is now a Ph.D. candidate in Computer Science at Virginia Commonwealth University. His research interests include Data Mining and Parallel Computing using GPU.

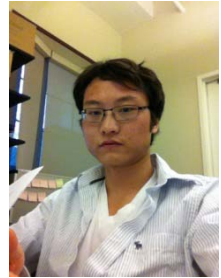

Robert Strack received his M. S. Eng. degree in Computer Science from AGH University of Science and Technology, Cracow, Poland in 2007. He is now working towards his Ph. D. degree in Computer Science at Virginia Commonwealth University, Richmond, US. His research is oriented towards Machine Learning and Data Mining algorithms and his field of interest includes Support Vector Machines classification and Parallel Computing.

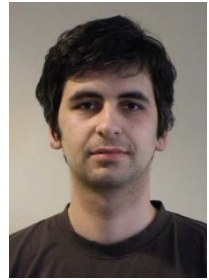

Erik Test is a Ph.D. student at Virginia Commonwealth University (VCU), Richmond, United States studying Computer Science and will complete his Master's degree in the May. 2011. He received his Bachelor's in Computer Science in 2007 from VCU. He also gained previous work experience at Acision, BAE Systems, and SENTEL. His research interests are high performance computing (HPC), GPU computing, and machine learning as well as machine learning in an HPC framework.

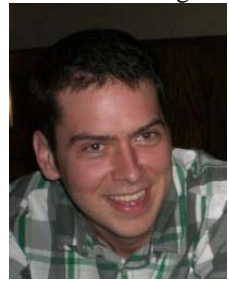